# Adhera: A Human-Centered Health Informatics Solution for Reducing Informal Caregiver Burden through Improved Medication Adherence

Zhiyin Zhou*
New York, United States

*Abstract*—The growing global population of older adults, combined with ongoing healthcare workforce shortages, has increased reliance on informal caregivers, including family members and friends who provide unpaid support to individuals with chronic illnesses. Among their daily responsibilities, medication management remains one of the most demanding and error-prone tasks. Non-adherence to prescribed regimens not only undermines patient outcomes but also intensifies caregiver stress, anxiety, and fatigue. Although digital health technologies have proliferated to address adherence, most solutions focus exclusively on patients and neglect the informational and emotional needs of caregivers.

This paper introduces Adhera, a caregiver-inclusive health informatics system designed to support medication adherence while reducing caregiver burden. Using a mixed-methods research design that included fifteen semi-structured caregiver interviews, sixty-five survey responses, and five pharmacist consultations, this study identified three primary challenges: caregiver stress related to uncertainty about medication intake, fragmented communication with healthcare professionals, and distrust in existing digital tools. Informed by the CeHRes Roadmap 2.0 and the Triple Bottom Line by Design and Culture (TBLD+C) framework, as well as recent co-design studies involving caregivers, Adhera integrates a sensor-equipped smart pill organizer with a mobile companion application that records intake events, sends real-time reminders, and provides caregivers with synchronized adherence data.

Preliminary evaluation suggests that Adhera enhances visibility, improves caregiver confidence, and streamlines medication routines. This study contributes to the field of health informatics by demonstrating how human-centered design and collaborative frameworks can align technical innovation with empathy-driven care.

## I. Introduction

Population aging is transforming healthcare systems worldwide. According to the United Nations, one in six people globally will be over the age of sixty-five by 2050 [1]. This demographic shift has accelerated the prevalence of chronic conditions such as diabetes, hypertension, and dementia, creating long-term demand for daily care management. Due to workforce shortages and the high cost of professional care, a large proportion of this responsibility now falls on informal caregivers, family members or friends who provide unpaid assistance [2], [5].

More than fifty-three million adults in the United States serve as informal caregivers, handling tasks such as medication administration, symptom tracking, and coordination with clinicians [2], [5]. Research has consistently documented the physical and psychological impact of caregiving, including high levels of stress, anxiety, and decreased quality of life [6], [7]. The Caregiver Identity Theory explains how caregiving evolves into a defining personal role that can produce role strain when caregivers balance personal needs with care obligations [4]. Recent evidence shows that this burden extends across multiple dimensions, encompassing emotional distress, cognitive fatigue, and reduced social participation [5], [23].

Medication management remains a critical component of caregiving. The World Health Organization reports that adherence to long-term therapy among patients with chronic conditions averages only fifty percent, resulting in treatment failure, hospital readmissions, and increased healthcare costs [3]. For caregivers, uncertainty about whether medications have been taken correctly is a major source of anxiety. Studies indicate that more than seventy percent of caregivers frequently worry about missed doses or incorrect administration [5], [9]. Non-adherence not only endangers patient health but also disrupts caregivers' emotional stability and sleep patterns [6], [10].

Digital health interventions have shown potential to alleviate some aspects of caregiver stress. For example, Zhai et al. (2023) demonstrated that mobile and telehealth interventions can strengthen caregiver confidence but rarely provide real-time verification of adherence [20]. Similarly, Zainal et al. (2025) identified usability, affordability, and digital literacy as barriers that limit the effectiveness of current digital tools for caregivers [21]. These findings align with Neller et al. (2024), who emphasized the need for caregiver-inclusive systems that recognize the emotional dimensions of care [27]. Despite rapid growth in mobile health applications and smart medication devices, most solutions remain patient-centric and do not allow caregivers to participate in monitoring or data sharing [8], [10], [13], [22], [28].

Existing digital health solutions, such as Dosecast, MyTherapy Pill Reminder, and Mango Health, primarily target patients who are technologically independent and capable of managing their own medication routines. These applications provide one-directional reminders and self-tracking features but do not enable shared access or real-time communication for caregivers. As a result, caregivers lack the visibility and reassurance needed to ensure adherence for patients who depend on them. Adhera addresses this unmet need by positioning caregivers as primary users. The system integrates a physical smart organizer that delivers reminders directly to patients with a synchronized mobile dashboard that allows caregivers to monitor adherence remotely, track refills, and share data with pharmacists. This caregiver-centered model introduces real-time collaboration and emotional reassurance that are largely absent from current adherence technologies.

To address this gap, this paper presents Adhera, a human-centered health informatics system that connects caregivers, patients, and healthcare professionals within a shared digital environment. Adhera builds upon the CeHRes Roadmap 2.0 framework for participatory eHealth design [15], the equity-centered design principles, and co-design best practices identified by Merchán-Baeza et al. (2023) and Parmar et al. (2025) [24], [26]. Together, these frameworks ensure that Adhera balances technical feasibility with emotional resonance and social inclusivity.

The guiding research question for this study is:

*How can human-centered design and health informatics approaches improve medication adherence and reduce caregiver burden among informal caregivers?*

The remainder of this paper is structured as follows: Section II reviews prior research on caregiver burden, medication adherence, and digital health technologies. Section III outlines the mixed-methods research design. Section IV presents the results and identified themes. Section V describes the Adhera system design and architecture. Section VI discusses theoretical and practical implications. Section VII concludes with contributions and recommendations for future work.

## II. RELATED WORK

### A. Informal Caregiver Burden

Informal caregivers play a critical role in supporting aging populations and individuals living with chronic diseases. They perform daily tasks such as medication organization, appointment scheduling, and symptom monitoring, often without formal training or compensation. Research consistently shows that sustained caregiving responsibilities are associated with psychological distress, physical fatigue, and diminished quality of life [5], [6], [7]. The Caregiver Identity Theory proposed by Montgomery and Kosloski [4] explains how caregiving responsibilities can gradually become central to a person's identity, often producing emotional strain when personal and caregiving roles conflict.

Recent scholarship has deepened understanding of these dynamics. Neller et al. (2024) demonstrated that caregiver stress fluctuates over the course of care and is heavily influenced by uncertainty and inadequate institutional support [27]. Zainal et al. (2025) found that caregivers face persistent barriers when engaging with digital health tools, including cost, limited digital literacy, and lack of cultural adaptation [21]. These findings confirm that caregiver burden extends beyond physical fatigue and includes emotional and cognitive stressors that current healthcare systems do not adequately address.

### B. Medication Adherence and Health Outcomes

Medication adherence remains one of the most challenging aspects of chronic disease management. The World Health Organization reports that only half of patients with long-term conditions follow their prescribed regimens, which results in disease progression, hospital readmissions, and higher mortality rates [3]. For caregivers, medication adherence is a continuous source of concern, as errors or omissions can have serious clinical consequences. Studies show that many caregivers experience anxiety, guilt, and disrupted sleep when uncertain whether a patient has taken their medication correctly [5], [9], [10].

The ABC taxonomy of adherence, developed by Vrijens et al. [11], categorizes adherence behavior into three stages: initiation, implementation, and discontinuation. This model recognizes adherence as a behavioral process influenced by time, environment, and human relationships. Although numerous interventions have been designed to improve adherence, systematic reviews indicate that their effects are often small and short-lived [12]. Zhai et al. (2023) noted that caregiver participation can enhance adherence rates but remains poorly integrated into most technology-based solutions [20]. These insights highlight the need for systems that incorporate caregivers as active participants rather than secondary observers

### C. Digital Adherence

Digital technologies have transformed medication adherence strategies over the past decade. Mobile applications, connected pill dispensers, and wearable sensors have improved tracking and reminders, yet their focus remains primarily on the patient experience [8], [9], [10], [13]. This limited perspective overlooks the informational and emotional needs of caregivers, who often serve as intermediaries between patients and healthcare providers.

A recent systematic review by Gargioni et al. (2024) surveyed existing smart medication devices and found that although most capture adherence data effectively, very few share this data directly with caregivers [22]. Lee et al. (2023) analyzed mobile health applications and concluded that current platforms emphasize unidirectional reminders instead of collaborative communication between caregivers and patients [23]. Likewise, Zainal et al. (2025) reported that even when caregivers are included, usability challenges and accessibility barriers frequently limit adoption [21].

From a technical perspective, IoT-based systems demonstrate promise in enabling remote monitoring. The Smart Pill Box project presented at the 2024 9th International Conference on Cloud Computing and Internet of Things described a web-connected pill organizer that allows clinicians to view medication intake in real time [28]. However, the system does not provide caregivers with direct feedback or interactive tools. Adhera advances this concept by offering a caregiver-facing dashboard, personalized alerts, and secure data synchronization, promoting shared accountability and emotional reassurance.

*D. Strategic Design and Human-Centered Informatics Integration*

Human-centered design and health informatics research increasingly converge on the principle that technological innovation must be socially responsive and participatory. The CeHRes Roadmap 2.0 advocates co-design processes that involve users at every stage of eHealth development to ensure functional relevance and sustained adoption [15]. Similarly, the Triple Bottom Line by Design and Culture (TBLD+C) framework integrates social, economic, and cultural dimensions into design practice, promoting equitable and sustainable, human-centered technology development [16].

Recent work has expanded on these principles in the context of caregiving. Merchán-Baeza et al. (2023) reviewed co-created technological interventions for caregivers and found that systems developed with caregiver input show greater long-term engagement [24]. Premanandan et al. (2024) proposed methodological guidelines for evaluating digital caregiver applications and emphasized ethical inclusion of user perspectives [25]. Parmar et al. (2025) demonstrated that collaborative design approaches can produce caregiver-centered systems that enhance both usability and emotional resonance [26]. Adhera's design approach draws from these frameworks by combining strategic design thinking with informatics engineering to create a solution that is technically effective, empathetic, and culturally adaptable.

*E. Research Gap*

Despite notable progress in digital health, most adherence systems remain narrowly patient-focused. While clinicians increasingly rely on connected technologies to track medication intake, caregivers are rarely granted equivalent access to adherence data or communication channels [13], [17], [18]. Studies by Zainal et al. (2025) and Lee et al. (2023) confirm that usability and digital literacy remain persistent barriers, especially among older caregivers [21], [23]. Gargioni et al. (2024) further noted that few systems include emotional reassurance or collaborative design features [22].

Adhera addresses this gap by reframing medication adherence as a shared ecosystem that connects patients, caregivers, and healthcare providers in real time. The system integrates IoT-enabled sensing, mobile communication, and participatory design to support behavioral adherence while reducing caregiver stress. By combining evidence-based informatics methods with empathic design, Adhera represents a step toward inclusive, equitable, and sustainable digital healthcare.

### III. METHODOLOGY

*A. Research Design*

This study employed a mixed-methods design that combined qualitative and quantitative data collection to understand caregivers' needs, behaviors, and expectations regarding medication management. The goal was to generate actionable insights that could directly inform the design of the Adhera system. This approach aligns with the CeHRes Roadmap 2.0 framework, which emphasizes user participation and iterative design throughout the development of digital health interventions [15].

The study was conducted in three sequential phases. The first phase explored caregiver challenges through in-depth interviews. The second phase gathered quantitative survey data to validate common pain points identified during the interviews. The third phase focused on collaborative design workshops to translate findings into concrete system requirements. This structure reflects best practices in co-created technology development for caregivers, as identified by Merchán-Baeza et al. (2023) and Parmar et al. (2025) [24], [26].

*B. Participants*

   *a) Caregiver Interviews*

A total of fifteen caregivers participated in semi-structured interviews, including ten informal caregivers (family members or friends providing unpaid support) and five professional caregivers employed by home health agencies. Participants ranged in age from 22 to 59 years and represented diverse cultural, socioeconomic, and caregiving backgrounds. The sample included adult children caring for aging parents, spouses supporting partners with chronic illness, and professional caregivers providing long-term assistance for elderly or disabled clients.

Caregiving duration varied substantially across participants. Four caregivers had been in their roles for less than two years, seven had between three and five years of experience, and four had provided care for more than a decade. Daily caregiving responsibilities ranged from three to five hours for most participants, although several full-time caregivers described continuous, day-long support routines. Approximately half of the care recipients were diagnosed with chronic medical conditions such as cancer, diabetes, or cardiovascular disease, while the remaining individuals required assistance due to Alzheimer's disease, dementia, age-related decline, or developmental disabilities.

   *b) Survey Respondents*

To complement the interviews, an online survey was distributed nationally using Google Forms. The survey received sixty-five valid responses, capturing caregivers aged 18 to 65 years. Approximately 68.3 percent of respondents identified as female, and over half (51.1 percent) cared for patients aged 54 years or older. Respondents supported individuals with a range of chronic conditions, including hypertension, diabetes, Alzheimer's disease, and cancer. Because the survey used convenience sampling, the results are not intended to be statistically representative of the national caregiver population, but they provide useful insight into common caregiving patterns and challenges.

   *c) Pharmacist Consultations*

Five pharmacists from both community and corporate pharmacy settings participated in one-on-one interviews. These consultations provided expert perspectives on medication adherence, patient education, and communication gaps between pharmacists, patients, and caregivers. Their input ensured that the Adhera design addressed practical concerns related to medication labeling, dosing schedules, and refill management.

*C. Data Collection Procedures*

Three complementary methods were used to collect data: semi-structured interviews, surveys, and participatory design workshops.

1. Semi-structured interviews: Fifteen one-hour interviews were conducted via video conferencing to ensure flexibility and accessibility. Each interview followed a guide based on themes from prior caregiving research [4], [5], [6], [7]. Questions focused on daily medication routines, sources of stress, coping strategies, and prior use of technology. Interviews were audio-recorded and transcribed verbatim for analysis.
2. Survey: A structured survey was distributed to sixty-five caregivers to quantify key challenges identified in the interviews. Survey items included Likert-scale questions on perceived stress, confidence in medication tracking, and prior experience with digital tools. This phase provided a broader understanding of the prevalence and intensity of the identified pain points.
3. Co-design workshops: Two participatory design sessions were held with small groups of caregivers to co-develop features for Adhera. Participants used scenario mapping and paper prototyping to envision how the system could support real-world medication routines. This method aligns with guidelines proposed by Premanandan et al. (2024), who emphasize participatory evaluation and ethical collaboration in caregiver-centered technology research [25]. Feedback gathered from these sessions directly informed Adhera's interface design and feature prioritization.

*D. Data Analysis*

The qualitative data from interviews and workshops were analyzed using thematic analysis to identify recurring patterns across participants. Coding was conducted manually by two researchers to ensure intercoder reliability. Themes were clustered into categories representing emotional, logistical, and informational challenges. This interpretive approach aligns with the Documentary Method, which focuses on how individuals make sense of lived experiences within social contexts [4].

Quantitative survey data were analyzed using descriptive statistics to measure frequency distributions and identify relationships between variables such as caregiver stress levels, number of medications managed, and prior technology use. Although inferential statistics were limited due to sample size, these quantitative patterns helped validate qualitative themes and supported data triangulation.

To ensure methodological transparency, a data-to-design mapping matrix was developed. This matrix linked each identified challenge (for example, "uncertainty about medication intake") to the specific design feature that addressed it (for example, "real-time adherence alerts"). This structure follows the participatory design logic recommended by Parmar et al. (2025), which emphasizes traceability between user insights and system features [26].

*E. Ethical Considerations*

The research protocol was reviewed by the university's institutional ethics committee and deemed exempt due to minimal risk. The researcher obtained informed consent from all participants before interviews and prototype testing. Participants were briefed on the study objectives, data handling procedures, and their right to withdraw at any time without penalty.

To protect participant privacy, all personal identifiers were removed from the interview transcripts and pseudonyms were used in all reporting. Research data were stored in encrypted form on a secure local drive with access limited to the researcher. The Adhera prototype followed a privacy-by-design approach. Device data were transmitted over Bluetooth Low Energy with its built-in encryption and then synchronized to the server using HTTPS secured through standard TLS protocols. Adherence data were stored in Firebase using encrypted storage services, and log files were processed locally to minimize external exposure of sensitive information.

During co-design workshops, participants were invited to express preferences regarding data visibility, consent, and information sharing. These procedures align with ethical standards for digital caregiver research established by Premanandan et al. (2024) [25] and with the equity-centered design principles outlined in the TBLD+C framework and the human-centered design approach proposed by Lyon et al. (2025) [16]. Collectively, these measures ensured that the study maintained confidentiality, transparency, and respect for participant autonomy throughout all phases of research and system testing.

*F. Validity and Reliability*

Several steps were taken to ensure data quality and credibility. Method triangulation across interviews, surveys, and workshops provided a comprehensive understanding of the caregiver experience. Member checking was performed by returning summary findings to a subset of participants for feedback and confirmation. Consistency in coding was verified by two independent researchers, yielding an intercoder agreement rate above eighty-five percent.

To strengthen the study's design reliability, the methodology adhered to iterative co-design principles, allowing feedback loops at multiple stages. This iterative process reflects the co-creation standards outlined in the CeHRes Roadmap 2.0 [15] and recent caregiver-centered studies that prioritize sustained stakeholder engagement throughout the development cycle [24], [26].

IV. RESULTS

The findings from interviews, surveys, and co-design sessions revealed three overarching themes that capture the primary challenges caregivers face in medication management: (A) emotional stress caused by uncertainty, (B) fragmented communication among caregivers, patients, and healthcare professionals, and (C) limited trust in existing digital tools. Each theme is described below, along with supporting evidence from both qualitative and quantitative data. All participant names used in this section are pseudonyms to protect confidentiality.

*A. Emotional Stress and Uncertainty*

Across all data sources, caregivers described emotional distress resulting from constant uncertainty about whether medications had been taken correctly. This concern was especially common among those managing multiple prescriptions or caring for patients with memory impairments. One caregiver described the experience as *"a daily guessing game that feels like a full-time job."* Survey data supported this sentiment, with 72 percent of respondents reporting that medication tracking was the most stressful aspect of caregiving. Prior studies similarly link uncertainty about medication intake to anxiety, sleep disruption, and psychological strain [5], [9], [10]. Neller et al. (2024) found that emotional fatigue increases when caregivers lack reliable feedback from the care recipient [27].

The interviews revealed that this uncertainty created a persistent sense of vigilance that intensified over time. Reflecting on this progression, one participant explained, *"I had no idea how bad it would get… it became like a constant around-the-clock situation"* (David). Sleep disturbance was common among caregivers managing night-time responsibilities. As one caregiver noted, *"I can't sleep at all… nurses would check in every night at midnight, and I had to take vitals at night too"* (Lauren). Others described how caregiving displaced their daily routines, with one participant saying, *"I feel like I don't have a life at all"* (Lauren). For caregivers managing complex or multiple care tasks, the pressure was described as overwhelming. One participant summarized this burden succinctly: *"Sometimes I wish I was like five people… one for each of them"* (Linda).

Co-design participants emphasized that real-time confirmation of medication intake would meaningfully reduce unnecessary worry and allow caregivers to concentrate on other responsibilities. This feedback directly informed the inclusion of instant adherence notifications and activity logs in Adhera's design.

*B. Fragmented Communication and Coordination*

Participants frequently expressed frustration with the lack of integrated communication between caregivers, healthcare professionals, and patients. Outdated or inconsistent medication lists, delayed updates, and unclear instructions often resulted in confusion and missed doses. Twelve of the fifteen interviewees reported that they rarely received timely updates from pharmacists or clinicians regarding medication changes. Pharmacists likewise indicated that caregivers often lacked access to essential prescription information. These findings align with existing literature documenting weak caregiver–clinician communication networks [17], [18], [19].

Caregivers also described limited coordination within family systems and among professional caregivers. One participant explained, *"There was no one that could substitute me. No rotations in our family"* (Lauren), illustrating a common reliance on a single individual without structured support. Professional caregivers noted similar challenges, including fragmented handoffs and an absence of communication across shifts. As one caregiver recounted, *"We also had no contact at all… we take turns with no speaking, we don't even see each other"* (Hui). Several caregivers emphasized that this lack of coordination made it difficult to manage the growing scope of their responsibilities, with one noting, *"I should have gotten help… it shouldn't be all on one person. You need staff"* (David).

Co-design workshop participants recommended a shared digital environment that would allow caregivers and healthcare professionals to access verified medication information and exchange notes. This input shaped Adhera's two-way data-sharing system, ensuring updates are synchronized across user dashboards and reducing information silos. This approach responds directly to gaps identified by Lee et al. (2023), who found that most mobile health applications rely on one-directional reminders rather than collaborative communication mechanisms [23].

*C. Distrust in Existing Digital Tools*

Although most caregivers were familiar with mobile applications and smart devices, they expressed skepticism about the reliability, usability, and security of existing adherence systems. Several participants described discontinuing the use of digital tools due to confusing interfaces, technical failures, or uncertainty about data accuracy. These observations align with findings by Zainal et al. (2025), who reported that caregivers frequently abandon digital tools due to usability challenges and limited cultural relevance [21], and by Gargioni et al. (2024), who found that many smart medication devices do not provide transparent data access or clear privacy assurances [22].

Participants also described broader distrust in care systems lacking transparency, which parallels concerns about digital solutions. One caregiver explained, *"You can never know 100% until you've worked with them… you can't be sure what is happening"* (Linda), highlighting the need for clear visibility into care activities. Another participant echoed this concern, stating, *"I have to trust they are doing the right thing, but I can't check as much anymore"* (Linda). Caregivers consistently emphasized the importance of reliability and confidence, with one remarking, *"I would prefer more supports that are reliable… something I can trust"* (Linda).

These concerns guided several design decisions in Adhera, including visible confirmation timestamps, simplified user interfaces, customizable data-sharing permissions, and flexible notification settings. Some caregivers preferred discrete alerts, while others favored dashboards accessible by multiple family members. This variation underscores the importance of configurable interfaces that adapt to different caregiving contexts.

*D. Quantitative Summary of Findings*

Table I summarizes the main quantitative findings from the caregiver survey. The results confirm that adherence-related uncertainty and poor communication are widespread, with over 70% of respondents reporting stress or sleep loss linked to medication management.

TABLE I. SUMMARY OF CAREGIVER-REPORTED CHALLENGES (N = 65)

| Category | Description | Percentage (%) |
|---|---|---|
| Patient forgets medication | Missed or late doses due to memory lapses | 79.4 |
| Incorrect dosage or timing | Confusion about medication labels and schedules | 55.6 |
| Communication difficulties | Limited ability to contact providers or pharmacists | 46.0 |
| Language or literacy barriers | Misunderstanding medical terminology | 40.0 |
| Use of digital reminders | Adoption of smartphone or pillbox reminder apps | 58.7 |
| Anxiety or sleep loss | Emotional stress linked to medication management | 72.3 |

*E. Design Implications*

The integration of findings from all data sources yielded three direct implications for design:

1. Caregiver visibility must be central. Systems should not only notify patients but also provide caregivers with real-time adherence confirmation and historical records.
2. Communication must be bidirectional. Digital adherence tools should allow for ongoing collaboration between caregivers and healthcare professionals rather than one-way updates.
3. Trust must be earned through transparency. Privacy controls, data accuracy, and interface simplicity are critical for sustained adoption.

These implications informed the development of the Adhera prototype, ensuring that the final system reflects caregivers' lived experiences and expectations. The participatory approach followed here corresponds with the principles of inclusive design emphasized in the CeHRes Roadmap 2.0 [15] and co-created technology frameworks outlined by Parmar et al. (2025) and Merchán-Baeza et al. (2023) [24], [26].

V. SYSTEM DESIGN: ADHERA

*A. Concept Overview*

The findings from the research informed the creation of Adhera, a caregiver-inclusive health informatics system that combines a smart medication organizer with a companion mobile application. The system is designed to reduce caregiver anxiety, increase adherence transparency, and promote collaborative medication management among caregivers, patients, and healthcare professionals.

Unlike conventional adherence tools that focus solely on the patient, Adhera treats both patients and caregivers as core users in the medication monitoring process. The system records each medication event, synchronizes the data securely in the cloud, and provides caregivers with visual confirmation in real time. This closed feedback loop transforms adherence into a shared task supported by data reliability and mutual trust.

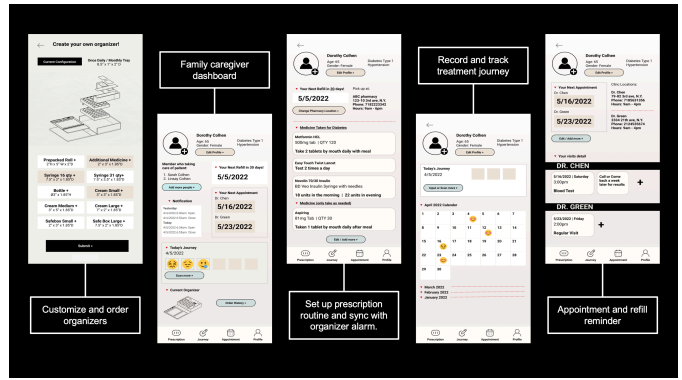

Fig. 1. Mobile Application Mock-Up of Adhera

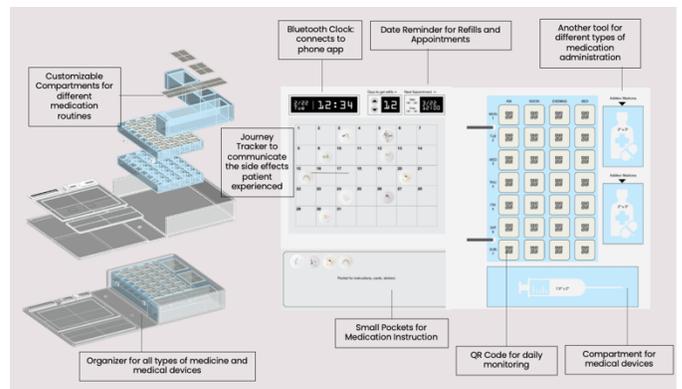

Fig. 2. Smart Pill Organizer Mock-Up of Adhera

*B. Design Objectives*

The Adhera system was developed with three primary design objectives, each derived from empirical insights gathered during the study and guided by the CeHRes Roadmap 2.0 [15] and Human-Centered Design framework [16].

1. Enhance visibility of adherence events: Provide caregivers with immediate confirmation when medication is taken to reduce uncertainty and emotional stress.
2. Reduce cognitive load: Simplify complex medication schedules through automation, visual cues, and multimodal alerts.
3. Facilitate shared responsibility: Enable caregivers, patients, and healthcare professionals to collaborate effectively through synchronized dashboards and transparent communication channels.

These goals reflect the participatory design principles advocated by Merchán-Baeza et al. (2023) and Parmar et al. (2025), who emphasize that caregiver participation during the design phase is essential for achieving empathy-driven and scalable outcomes [24], [26].

## C. System Architecture

Adhera follows a modular Internet of Things (IoT) architecture comprising three interconnected layers:

1. Hardware Layer – Smart Organizer: The organizer includes infrared and weight sensors embedded within each compartment to detect when a pill is removed. A low-power microcontroller and Bluetooth Low Energy (BLE) module transmit timestamped data to the mobile app. Each compartment is color-coded by time of day, and the device operates using a rechargeable lithium-ion battery. The ergonomic design supports one-handed use for individuals with limited dexterity.

2. Software Layer – Mobile Application: The companion mobile app, developed using Flutter for cross-platform compatibility, serves as the user interface for both caregivers and patients. The app receives real-time sensor data, displays adherence events, and sends reminders via sound, vibration, and on-screen notifications. Accessibility features include large font options, high-contrast text, and multilingual support in English, Spanish, and Mandarin, reflecting cultural adaptability recommended by Zainal et al. (2025) [21].

3. Cloud Layer – Data Management and Analytics: The adherence data are transmitted to Google Firebase Cloud, where they are encrypted and stored using Advanced Encryption Standard (AES) 256-bit encryption. Secure communication between devices and the server uses Transport Layer Security (TLS) 1.3 to prevent unauthorized access. The cloud engine processes adherence trends, generates summary reports, and enables data visualization through the caregiver dashboard. Adhera's API supports HL7 FHIR protocol integration for potential electronic health record (EHR) interoperability [15], [22].

## D. User Interaction Flow

Adhera's workflow was designed to align naturally with caregivers' daily routines, minimizing manual data entry. The process includes five key steps:

1. Setup: The caregiver or pharmacist enters the medication details, including name, dosage, and schedule, into the app. The system configures reminders automatically.
2. Notification: At the scheduled time, the smart organizer lights up and emits a short sound, while the mobile app sends a reminder.
3. Medication Intake: When the patient opens the compartment, the sensor records the event and timestamps it.
4. Data Synchronization: The event is uploaded to the cloud, and caregivers receive an instant notification confirming adherence.
5. Review: Caregivers and healthcare professionals can view adherence summaries and trend reports through their dashboards or data exports.

This flow reflects the participatory co-design outcomes described in the literature, where simplicity and confirmation feedback were prioritized to reduce anxiety and foster trust [24], [26]. It also enables caregivers to monitor adherence non-intrusively, balancing autonomy for the patient with reassurance for the caregiver.

## E. Prototype Evaluation

A low-fidelity prototype of Adhera was tested with five caregivers from the interview cohort over a seventy-two-hour simulation of typical medication routines. The evaluation focused on usability, perceived value, and functional reliability. All participants reported that real-time adherence confirmation reduced anxiety and improved confidence in managing medication events. Most found the interface intuitive and favored the combined use of visual and auditory alerts.

Participants suggested enhancements such as customizable alert tones, multilingual prompts, and battery-status notifications. Overall, the prototype performed reliably across the testing period, demonstrating stable synchronization and consistent data transmission. These findings are consistent with prior studies indicating that IoT-based adherence tools can improve user trust and perceived control [21], [28]. Feedback from this phase guided the development of the high-fidelity prototype, with emphasis on accessibility and sustained reliability.

## F. Privacy and Data Security

Adhera's modular design supports flexible deployment in both household and clinical settings. The hardware uses low-cost components, and the software architecture is built on widely adopted open-source frameworks for mobile development, data synchronization, and encryption, which reduces development and maintenance expenses.

Three potential implementation pathways are envisioned:

1. Consumer Use: Families purchase Adhera for home medication management.
2. Clinical Integration: Hospitals recommend Adhera for post-discharge monitoring to prevent readmissions.
3. Pharmacy Partnerships: Pharmacies offer Adhera as a service bundle that connects dispensing data to adherence tracking.

This scalability aligns with the sustainable innovation principles described in the HCD framework [16] and supports equitable access to digital health tools.

## VI. DISCUSSION

### A. Integrating Caregiver Needs into Health Informatics

The results demonstrate that informal caregivers are essential participants in medication adherence and should be treated as co-users rather than peripheral observers. Although many adherence tools exist, they typically focus on patient self-management and rarely address caregiver uncertainty or emotional distress. Zhai et al. (2023) and Zainal et al. (2025) have emphasized that digital health interventions often fail to meet caregiver needs due to limited feedback, lack of collaboration tools, and usability barriers [20], [21].

Adhera responds to these gaps by integrating real-time confirmation and caregiver-facing dashboards into the adherence workflow. The system provides continuous visibility of medication events, reducing anxiety and enabling caregivers to plan their daily activities more confidently. By reframing caregivers as active participants in the adherence process, Adhera aligns with emerging research that highlights the importance of family-centered digital health systems [2], [26].

This inclusion supports the concept of distributed care, where caregiving responsibilities are shared across households, clinics, and pharmacies. It also enhances the overall reliability of adherence data, as caregivers can confirm and cross-check medication behaviors directly through the system interface.

*B. Application of Human-Centered and Strategic Design Frameworks*

The design process for Adhera demonstrates how strategic and human-centered frameworks can shape technological innovation in healthcare. Guided by the Triple Bottom Line by Design and Culture (TBLD+C) framework, the system incorporates social, economic, and cultural considerations that ensure equitable access and long-term sustainability. The CeHRes Roadmap 2.0 further informed the participatory design process by structuring user involvement throughout each stage of development [15].

Studies by Merchán-Baeza et al. (2023), Premanandan et al. (2024), and Parmar et al. (2025) show that caregiver-inclusive co-design leads to systems that are more empathetic, usable, and accepted in real-world settings [24], [25], [26]. These frameworks shaped decisions regarding accessibility features, privacy controls, and visual simplicity. The emphasis on empathy and collaboration distinguishes Adhera from traditional IoT adherence systems, which often prioritize technical efficiency over human experience [22], [28].

By embedding caregiver voices directly into the design process, Adhera contributes to a broader shift in digital health design philosophy, demonstrating that emotional and relational factors can coexist with technical rigor.

*C. Implications for Healthcare Practice*

The study's findings suggest multiple pathways for integrating Adhera into existing healthcare infrastructure.

1. Continuity of care: Hospitals can implement Adhera as part of discharge planning for patients with chronic conditions. The caregiver dashboard would allow families to monitor adherence remotely and identify early signs of noncompliance, reducing the likelihood of readmission.

2. Professional engagement: Healthcare professionals can use Adhera's shared data interface to review adherence histories and provide targeted counseling. This process aligns with recent work by Ilardo and Speciale (2020), which highlights the potential for healthcare professionals to improve communication and trust with patients [14].

3. Community and home-based care: Organizations that support aging-in-place initiatives can distribute Adhera as a low-cost solution for managing medication adherence at scale.

These applications support a more collaborative model of care where information flows seamlessly between caregivers, patients, and healthcare professionals. Such connectivity aligns with the principles of integrated care described by Wolff et al. (2020) and Coleman and Roman (2015) [17], [18].

*D. Theoretical Implications*

From a theoretical standpoint, this study advances the discourse on caregiver-inclusive health informatics by operationalizing caregiver participation within system architecture. Building on the Caregiver Identity Theory [4], Adhera helps mitigate role strain by validating the caregiver's contribution through transparent and accessible data. This digital affirmation reduces feelings of helplessness and reinforces the caregiver's sense of competence.

Furthermore, by connecting strategic design frameworks with behavioral adherence models such as the ABC taxonomy [11], the study demonstrates that adherence is not solely a technical issue but a relational process shaped by trust and accountability. The integration of empathy and data-driven insights positions Adhera within a growing body of research that seeks to balance functionality with emotional well-being in digital health [16], [24], [26].

*E. Limitations*

Several limitations should be acknowledged. The study involved a relatively small sample size, which limits the generalizability of the findings. The prototype testing period was brief, preventing assessment of long-term behavioral changes or sustained system use. Future evaluations should include larger and more diverse participant groups across different caregiving contexts to validate outcomes statistically.

Additionally, although the prototype successfully demonstrated technical feasibility, adherence data were collected in simulated conditions rather than real-world environments. Future trials should include continuous data logging to evaluate long-term performance and system reliability. The study also relied on self-reported measures of caregiver stress, which could be complemented by validated psychological scales in future research.

Finally, while Adhera's current design supports multiple languages, additional localization and cultural testing are necessary to ensure global applicability, as suggested by Zainal et al. (2025) [21].

*F. Future Work*

Future research will expand the scope of Adhera through both technical and clinical validation.

1. Pilot Study: A three-month field trial with at least fifty caregiver–patient pairs will evaluate adherence accuracy, stress reduction, and usability metrics such as the System Usability Scale (SUS).

2. EHR Integration: The next iteration will integrate Adhera's API with electronic health records using HL7 FHIR protocols to allow clinicians to access adherence data securely.

3. Machine Learning Analytics: Predictive algorithms will be incorporated to identify early signs of non-adherence or caregiver burnout.
4. Policy and Implementation: Collaborations with healthcare payers and policymakers will explore reimbursement models and inclusion in chronic disease management programs.

This roadmap aims to position Adhera as a scalable and sustainable digital health solution that bridges technological innovation with compassionate caregiving.

*G. Contribution to Health Informatics Research*

Adhera advances the field of health informatics by introducing a caregiver-inclusive model that merges emotional intelligence with technical precision. The system illustrates how co-design principles can be translated into tangible innovations that improve both clinical outcomes and quality of life. By integrating behavioral, emotional, and technical insights, this research expands the definition of success in digital health to include psychological well-being, shared accountability, and user trust.

In doing so, Adhera contributes to a more holistic vision of healthcare where technology not only manages data but also nurtures human connection, empathy, and collective responsibility.

## VII. CONCLUSION

This study presented Adhera, a caregiver-inclusive health informatics system designed to improve medication adherence and alleviate the emotional burden experienced by informal caregivers. The mixed-methods research approach identified three primary challenges in caregiving: emotional stress caused by medication uncertainty, fragmented communication with healthcare professionals, and distrust in existing digital tools. By integrating findings from qualitative interviews, quantitative surveys, and co-design sessions, the research translated real-world caregiver needs into tangible system features.

Adhera combines a smart medication organizer with a connected mobile application that records intake events, issues timely reminders, and synchronizes adherence data with caregiver dashboards. The system reframes medication adherence as a shared process that connects patients, caregivers, and healthcare professionals within a transparent digital ecosystem. Guided by frameworks such as the CeHRes Roadmap 2.0 [15] and TBLD+C, Adhera demonstrates how empathy, equity, and usability can coexist with technical rigor in digital health innovation.

Preliminary evaluation results show that Adhera effectively reduces caregiver anxiety, improves perceived control, and simplifies medication management routines. The system's modular design and open-source foundation enable broad scalability in both home and clinical settings. These attributes align with recent findings by Merchán-Baeza et al. (2023) and Parmar et al. (2025), which highlight the long-term benefits of co-created caregiver technologies [24], [26].

Beyond its functional contributions, this research advances the theoretical understanding of caregiver-inclusive informatics. By validating the caregiver's role as a co-user and co-manager of adherence data, Adhera operationalizes the principles of Caregiver Identity Theory [4] within a digital framework. The system thus bridges emotional well-being and data transparency, illustrating how technology can enhance both relational trust and clinical reliability.

Future work will focus on extended field trials, machine learning integration for predictive adherence analytics, and secure interoperability with electronic health records. These directions aim to position Adhera as a scalable, ethical, and sustainable model for next-generation health informatics.

In conclusion, Adhera contributes to a growing movement in healthcare technology that seeks to balance precision with compassion. By embedding empathy, collaboration, and accountability within the structure of a digital system, Adhera exemplifies how human-centered informatics can support not only patient outcomes but also the emotional health and dignity of those who care for them.